\documentclass[aps,twocolumn,prb,showpacs,superscriptaddress]{revtex4}
\usepackage{graphics,bm}
\usepackage{amssymb}
\usepackage{epsfig}
\usepackage{epsf}

\def\be{\begin{equation}} \def\ee{\end{equation}}
\def\bea{\begin{eqnarray}} \def\eea{\end{eqnarray}}

\def\nn{\nonumber}

\begin{document}

\title{Impact of  Dynamic Orbital Correlations on Magnetic Excitations in
  the Normal State of Iron-Based Superconductors}

\author{Wei-Cheng Lee}
\email{leewc@illinois.edu}
\affiliation{Department of Physics, University of Illinois, 1110 West Green Street, Urbana, Illinois 61801, USA}

\author{Weicheng Lv}
\affiliation{Department of Physics, University of Illinois, 1110 West Green Street, Urbana, Illinois 61801, USA}

\author{J. M. Tranquada}
\affiliation{Condensed Matter Physics and Materials Science Department, Brookhaven National Laboratory, Upton, New York 11973, USA}

\author{Philip W. Phillips}
\email{dimer@vfemmes.physics.illinois.edu}
\affiliation{Department of Physics, University of Illinois, 1110 West Green Street, Urbana, Illinois 61801, USA}

\date{\today}

\begin{abstract}
We show here that orbital degrees of freedom produce a distinct signature
in the magnetic excitation spectrum of iron-based superconductors
above the magnetic ordering temperature.
Because $d_{xz}$ and $d_{yz}$ orbitals are strongly connected with Fermi surface 
topology, the nature of magnetic excitations can be modified
significantly due to the presence of either static or fluctuating orbital correlations.
Within a five-orbital itinerant model, we show that static orbital order generally leads to an enhancement
of commensurate magnetic excitations even when the original Fermi
surface lacks nesting at $(\pi,0)$ or 
$(0,\pi)$.  When long-range orbital order is absent, Gaussian
fluctuations beyond the standard random-phase approximation capture the effects of fluctuating orbital correlations on the magnetic excitations.
We find that commensurate magnetic excitations can also be enhanced if the orbital correlations are strong.
Our results offer a natural explanation for the incommensurate-to-commensurate transformation observed in a recent neutron scattering 
measurement (Z. Xu, et. al., arXiv:1201.4404), and we propose that this unusual transformation is an important signature to distinguish 
orbital from spin physics in the normal state of pnictides. Implications for the magnetic and superconducting states are discussed.
\end{abstract}

\maketitle

\section{Introduction}

The orbital-dependence of Fermi surface pockets due to the partial occupancy of multiple d-states of iron is a key
microscopic feature which distinguishes iron-based
superconductors from cuprates. It is natural, therefore, that both orbital and spin-based scenarios have been
advanced to account for the structural and magnetic transitions in
these materials. 
In the orbital-based scenario\cite{lv2009,kruger2009,lee_cc2009,chen_cc2010,lv2010,nevidomskyy2011}, the structural phase transition is induced by a spontaneous orbital order 
in the quasi-one-dimensional $d_{xz}$ and $d_{yz}$ orbitals which
breaks the $C_4$ symmetry. Such order induces a stripe-like
antiferromagnetic (AFM) transition at a lower temperature. 

In contrast, some have proposed that a spin-only model accounts for
both structural and magnetic transitions in pnictides.
While the source of the interactions that drive the magnetism differ
in the strong-coupling\cite{si2008,yildirim2008,xu2008,chen_f2008} and
weak-coupling\cite{qi2009,fernandes2010,fernandes2012} models
proposed,  both spin-based scenarios eventually end up with the 
same effective theory\cite{fernandes2012} with a magnetic order parameter lying in the 
manifold of $Z_2\times O(3)$. The structural phase transition is
interpreted as a breaking of the $Z_2$ symmetry, and the subsequent stripe-like AFM transition breaks the $O(3)$ symmetry.
The main difficulty in distinguishing between these two scenarios is that
both lead to the 
same Ginzburg-Landau effective action.

Although intensive research efforts has focused on the spin excitation spectrum in stripe-like AFM\cite{brydon2009,knolle2010,kaneshita2010}
and superconducting\cite{korshunov2008,graser2009,kemper2010} states,
a theoretical study on the effect of orbital order and fluctuations on
the spin excitation spectrum in the paramagnetic normal state is far
from complete.  It is this problem that we tackle here.  Our line of reasoning is based on the fact that 
if the spin interaction is the only energy scale driving both phase
transitions, the difference in magnetic excitations between the normal
and ordered states should be tied solely to a redistribution of
spectral weight. The positions of the peaks in the Brillouin zone,
namely the magnetic spectrum, should not undergo any qualitative change.
On the other hand, if there is an additional energy scale playing an
equally important role, then this extra energy scale could result in
qualitatively distinct changes in the magnetic spectrum above and
below the ordering transition.  

In this paper, we analyze the magnetic excitations within a five-orbital itinerant model with generalized on-site Hubbard interactions. 
and superconducting\cite{korshunov2008,graser2009,kemper2010} states, 
Since our main purpose is to identify several features which can not be captured without involving orbital degrees of freedom,
an orbital-dependent non-spin interaction is introduced in order to investigate systematically how orbital correlations affect the magnetic 
excitations in the paramagnetic normal state. 
We find that orbital order always enhances one of the commensurate spin excitation wave vectors even when the original Fermi surfaces lack nesting
at these commensurate wave vectors.
Moreover, to include the effect of fluctuating orbital correlations on the magnetic excitations, 
we derive a Gaussian fluctuation model to go beyond the standard random-phase approximation (RPA) treatment. 
Even in this case, we find that fluctuating orbital correlations also tend to shift the spectral weight toward the commensurate wave vector, resulting in a
broad spectrum near the commensurate wave vectors as seen in neutron scattering measurements.
We show that our results offer a natural explanation for the incommensurate-to-commensurate transformation observed in the inelastic neutron scattering measurement on
Fe$_{1-x}$Ni$_x$Te$_{0.5}$Se$_{0.5}$\cite{xu2012}, which can hardly be
explained by spin-based scenarios.  As there is no reason to suspect
that one model is insufficient to explain the normal state of the
pnictides, we conclude that orbital physics is the key player in
driving both the structural as well as the magnetic transitions in the
iron-based superconductors. 
We propose further experimental studies on magnetic excitations
at higher temperature in the normal state to further pin down the
orbital scenario.

\section{Formalism}
\subsection{Random phase approximation (RPA) theory for magnetic excitation}
Starting from a tight-binding model in the unfolded Brillouin zone with one Fe atom per unit cell, we write the model Hamiltonian as
\bea
H&=&H_t + H_{OO} + H_I \nn\\
H_I &=& \sum_{ia} U n_{ia\uparrow}n_{ia\downarrow} + \sum_{i,b>a}(U'-\frac{J}{2}) n_{ia} n_{ib}\nn\\
 &-& \sum_{i,b>a}2J\vec{S}_{ia}\cdot \vec{S}_{ib} + J'\big(p_{ia}p_{ib}^\dagger + h.c.\big),
\label{hamI}
\eea
where
\bea
n_{ia} \equiv \sum_{\sigma} c^\dagger_{ia\sigma}c_{ia\sigma},\nn\\
\vec{S}_{ia}\equiv c^\dagger_{ia\mu}\vec{\sigma}_{\mu,nu}c_{ia\sigma\nu},\nn\\
p_{ia} \equiv c_{ia\downarrow}c_{ia\uparrow}.
\eea
We adopt the relationships $U'=U-2J$, $J=J'$ and throughout the paper we use the values of $U=2.0$eV and $J=0.2$eV.
$H_t$ is the five-orbital tight-binding model fitted by Graser {\it et. al.}\cite{graser2009}, and $H_{OO}$ is the effective interaction for `orbital order' 
which we put in by hand as
\bea
H_{OO} &=& \frac{1}{N}\sum_{\vec{q}} \eta_q \zeta_{-\vec{q}}\zeta_{\vec{q}}\nn\\
\zeta_{\vec{q}}&=&\sum_{\vec{k},\sigma} 
\big(c^\dagger_{xz,\sigma}(\vec{k}+\vec{q}) c_{xz,\sigma}(\vec{k}) - c^\dagger_{yz,\sigma}(\vec{k}+\vec{q}) c_{yz,\sigma}(\vec{k})\big).\nn\\
\eea
We do not justify the microscopic origin of $H_{OO}$ since the
conclusion is independent of the mechanism for orbital order.
However, we will comment on the differences in the magnetic excitations for different scenarios in the discussion section.
For ferro-orbital order, we have $\eta_q=-\eta_0$ for $\vec{q}=0$,
which leads to a mean-field Hamiltonian of the form,
\be
H^{MF} = H_t - \phi\,\sum_{\vec{k},\sigma} 
\big(c^\dagger_{xz,\sigma}(\vec{k}) c_{xz,\sigma}(\vec{k}) - c^\dagger_{yz,\sigma}(\vec{k}) c_{yz,\sigma}(\vec{k})\big)
\ee
with $\phi = \frac{\eta_0}{N}\langle \zeta_{\vec{q}=0}\rangle$ solved self-consistently.
It is convenient to use the basis,
$\psi_{\vec{k},\sigma}^\dagger\equiv \big(c^\dagger_{\vec{k},xz,\sigma},c^\dagger_{\vec{k},yz,\sigma},c^\dagger_{\vec{k},xy,\sigma},
c^\dagger_{\vec{k},x^2-y^2,\sigma},c^\dagger_{\vec{k},3z^2-r^2,\sigma}\big)$, and introduce $\hat{U}^\sigma_{\vec{k}}$ such that
$\big(\hat{U}^\sigma_{\vec{k}}\big)^\dagger \hat{H}^{MF}(\vec{k},\sigma) \hat{U}^\sigma_{\vec{k}}
= {\rm diag.}\big[E_{\vec{k},1,\sigma},\cdots,E_{\vec{k},5,\sigma}\big]$.
It then follows that the bare response function can be expressed as
\bea
&&\chi^{(0)}_{ab;cd}(\vec{q},i\omega_n,\phi)\nn\\
&=&-\frac{1}{N}\sum_{\vec{k},\sigma,l,m} 
\big(\hat{U}^\sigma_{\vec{k}+\vec{q}}\big)_{a,l}\big(\hat{U}^\sigma_{\vec{k}+\vec{q}}\big)^*_{c,l}
\big(\hat{U}^\sigma_{\vec{k}}\big)_{d,m}\big(\hat{U}^\sigma_{\vec{k}}\big)^*_{b,m}\nn\\
&\times&\frac{n_F(E_{\vec{k}+\vec{q},l,\sigma}) - n_F(E_{\vec{k},m,\sigma})}{E_{\vec{k}+\vec{q},l,\sigma} - E_{\vec{k},m,\sigma} - i\omega_n}.
\eea
Using the same notation as Graser {\it et. al.}\cite{graser2009}, we
express the spin response function to one-loop as
\be
\chi^{RPA}(\vec{q},\omega,\phi) = \chi^{(0)}(\vec{q},\omega,\phi)\big[1-\hat{V}^s\chi^{(0)}(\vec{q},\omega,\phi)\big]^{-1}
\ee
where $\hat{V}^s$ is the spin interaction kernel derived in Ref.[\onlinecite{kemper2010}].

\subsection{Gaussian fluctuations in the absence of long-range order}

Even above the transition for long-ranged orbital order, the magnetic
excitations can still be influenced by the fluctuations in the
proximity to the orbitally ordered state. 
In this section, the minimal Gaussian fluctuation model is derived from the path integral formalism.
Since our focus is on the interplay between spin-flip excitations and orbital ordering in the normal state without any long-range order,
we introduce the Hubbard-Stratonovich fields to decouple $H_{OO}$ and the spin-flip
channels in $H_I$ 
\be
S^+_{ab}(\vec{q})\equiv \sum_{\vec{k}} c^\dagger_{\vec{k}+\vec{q},a,\uparrow}c_{\vec{k},b,\downarrow},
\ee
and drop all the other terms.  The partition function takes the form,
\bea
Z&=&\int D[c]D[c^\dagger] e^{-S_0-S_I-S_{OO}}\nn\\
&\propto& \int D[\phi]D[\phi^*] D[M]D[M^*] D[c]D[c^\dagger] e^{-S_0-S_M+S'_{OO}}\nn\\
S_{M} &=& \sum_{\vec{k},\vec{q},i\omega_n} \sum_{abcd} \big[M_{ab}(\vec{q},i\omega_n) S^-_{cd}(\vec{q},i\omega_n)\nn\\
&+& M_{ab}(-\vec{q},-i\omega_n)S^+_{cd}(\vec{q},i\omega_n)\nn\\
&+&[\hat{V}^s]^{-1}_{ab,cd}(\vec{q},i\omega_n)M_{ab}(-\vec{q},-i\omega_n) M_{cd}(\vec{q},i\omega_n)\big]\nn\\
S'_{OO} &=& \sum_{\vec{q},i\omega_n} \big[\phi(\vec{q},i\omega_n) \zeta(-\vec{q},-i\omega_n) + \phi(-\vec{q},-i\omega_n) \zeta(\vec{q},i\omega_n) \nn\\
&&+\eta_q^{-1}\phi^*(-\vec{q},-i\omega_n) \phi(\vec{q},i\omega_n)\big].
\label{partition}
\eea
Because we are interested in the case where the system is on the verge
of ferro-orbital order but not close to any instability in spin channel,
the largest weight in the partition function comes from the Gaussian fluctuations around $(\vec{q}=0,i\omega_n=0)$ in the $\phi$ field.
Therefore, we keep only the $\phi(\vec{q}=0,i\omega_n=0)$ field in Eq. \ref{partition} and treat the spin part by the standard saddle point approximation.
After integrating out the fermionic fields and expanding all the
Hubbard-Stratonovich fields up to quadratic order, we arrive at a
partition function of the form,
\bea
Z&\approx& \int D[\phi(0)]D[\phi(0)^*] D[M]D[M^*] e^{-S_{m} - S_{oo}}\nn\\
S_{m} &=& \sum_{\vec{k},\vec{q},i\omega_n} [\chi^{RPA}]^{-1}_{ab,cd}(\vec{q},i\omega_n,\phi(0))\nn\\
&&\times M_{ab}(-\vec{q},-i\omega_n) M_{cd}(\vec{q},i\omega_n)\big]\nn\\
S_{oo} &=& \lambda^{-2} \phi^*(0)\phi(0),
\eea
where $\lambda^{-2}=\eta_0^{-1} - \Pi(\vec{q}=0,i\omega=0)$ is the stiffness for $\phi(0)$ field renormalized by the bubble diagrams.
Then it is straightforward to arrive at the spin response function with minimal Gaussian fluctuations:
\be
\chi^{Gaussian}(\vec{q},\omega) = \frac{\int d\phi e^{-\phi^2/\lambda^2} \chi^{RPA}(\vec{q},\omega,\phi)}{\int d\phi e^{-\phi^2/\lambda^2}}
\label{chigaussian}
\ee
where $\lambda^2$ is an effective parameter measuring the strength of the fluctuating orbital correlations.
The physical meaning of the above Gaussian fluctuation model is clear. Because the system is on the verge of ferro-orbital order, the fluctuations of the $\phi$ fields
at $(\vec{q}=0,i\omega_n=0)$ are dominant. Either quantum or thermal fluctuations can create a temporary ferro-orbital order with a probability of $e^{-\phi(0)^2/\lambda^2}$, 
which is the leading fluctuating orbital correlations in this case.
It is worth mentioning that the orbital order parameter $\langle \phi\rangle=0$ is strictly fulfilled, so there is no long-range order.
This is analogous to the formalism used by Gollub {\it et. al.} to study diamagnetism above ${T}_{c}$ in conventional superconductors\cite{gollub1973}. 

\begin{figure}
\includegraphics{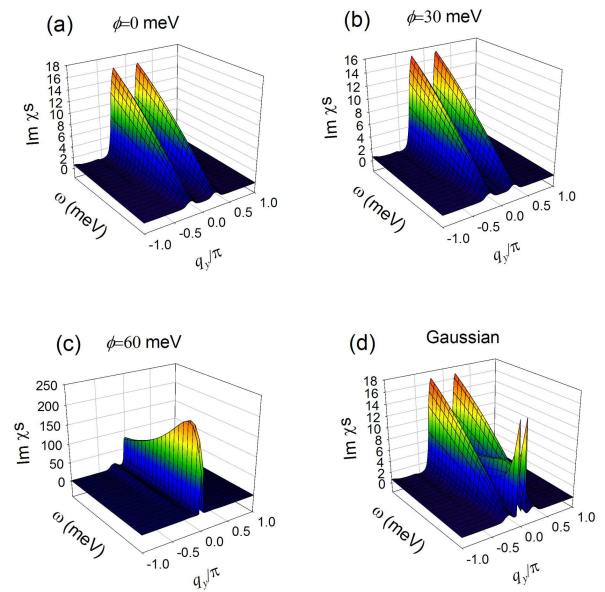}
\caption{\label{fig:incomm-comm} The imaginary part of the spin susceptibility are plotted along $\vec{q}=(\pi,q_y)$ for (a) $\phi=0$, (b) $\phi=30$ meV, (c) $\phi=60$ meV, and (d) Gaussian fluctuations with $\lambda^2=60$ meV. In this figure, the chemical potential $\mu$ is set to be 0 in the tight-binding model proposed in 
Ref.[\onlinecite{graser2009}], corresponding to an electron doping of about $5\%$.}
\end{figure}

\section{Results}
\begin{figure}
\includegraphics{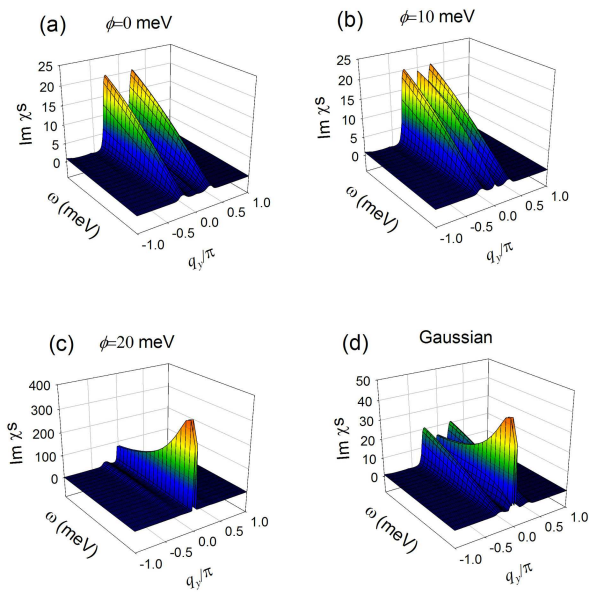}
\caption{\label{fig:incomm-comm-2} The imaginary part of the spin susceptibility are plotted for doping level about $9\%$ for (a) $\phi=0$, (b) $\phi=10$ meV, (c) $\phi=20$ meV, and (d) Gaussian fluctuations with $\lambda^2=30$ meV.}
\end{figure}

Our main results are summarized in Figs. \ref{fig:incomm-comm}, \ref{fig:incomm-comm-2}, and \ref{fig:demo}. 
The imaginary part of the spin susceptibility is plotted along the direction transvers to the $\vec{Q}_{AFM}=(\pi,0)$ for different cases. 
From Figs. \ref{fig:incomm-comm}(a)-(c), it can be seen clearly that with the increase of the 
orbital order parameter, the peaks in the magnetic excitation gradually move from incommensurate to commensurate wave-vectors, which can be understood solely in terms of 
the Fermi surface topology. 
Generally speaking, the realistic Fermi surfaces of the iron-based superconductors does not have perfect nesting at $(\pi,0)$ and $(0,\pi)$, and previous calculations 
have shown that in the normal state, the peaks in the spin susceptibility are usually incommensurate away 
from $(\pi,0)$, $(0,\pi)$\cite{kuroki2008,graser2009,kaneshita2010,ewings2011}.
However, due to the presence of orbital order, the Fermi surfaces are distorted along one of the in-plane axes, which always helps the enhancement of the magnetic excitations 
at $(\pi,0)$ for $\phi>0$ (or $(0,\pi)$ for $\phi<0$). It follows then
that the stripe-like antiferromagnetism accompanying 
pre-existing orbital order should have $\vec{Q}_{AFM}=(\pi,0)$ for $\phi > 0$ or $\vec{Q}_{AFM}=(0,\pi)$ for $\phi < 0$, which has been confirmed by previous 
mean-field \cite{lv2011} and first principle calculations\cite{lee_cc2009}.

The magnetic excitations are also strongly modified without long-range orbital order but with fluctuating orbital correlations. 
In this case, because the fluctuating orbital correlations induce fluctuations of the Fermi surface shape, the magnetic excitations at the incommensurate wave vectors 
are significantly diffused toward the commensurate ones. 
This results in a broadened spectrum near $(\pi,0)$, which can be seen from the results of our Gaussian fluctuation model 
shown in Fig. \ref{fig:incomm-comm}(d).
The imaginary part of the spin susceptibility are plotted for different doping level in Fig. \ref{fig:incomm-comm-2}, which shows the same behavior.

\begin{figure}
\includegraphics{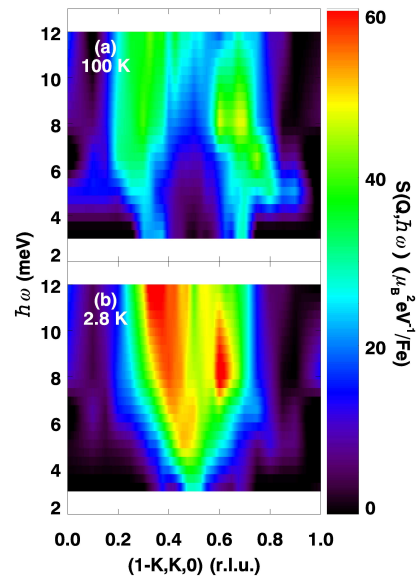}
\caption{\label{fig:fete-ins} Inelastic neutron scattering measurement (data from Xu {\it et. al.}\cite{xu2012}) of Fe$_{1-x}$Ni$_x$Te$_{0.5}$Se$_{0.5}$ with $x=0.04$ along the transverse direction to $Q_{AFM}$ for (a) high temperature normal state ($T>T_{fluc}$) and (b) the superconducting state ($T<T_c$).
An incommensurate-to-conmmensurate transformation is observed as the sample is cooled down.}
\end{figure}

\begin{figure}
\includegraphics{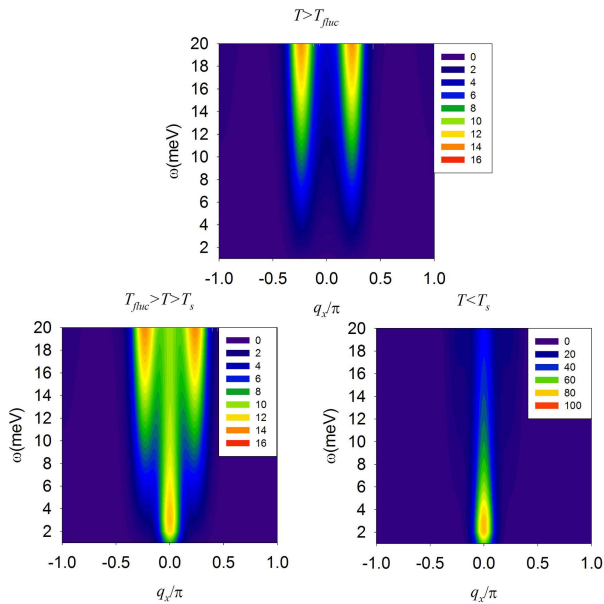}
\caption{\label{fig:demo} Demonstration of the evolution of the imaginary part of the spin susceptibility along $\vec{q}=(\pi,q_y)$ as the temperature is lowered from
high temperature. The parameters in the tight-binding model are the same as Fig. \ref{fig:incomm-comm}, and a broadening in the $\vec{q}$-space of $q_0 =0.1\pi$ 
is introduced with the form of ${\rm Im}\chi^{B}(\vec{q},\omega) = \sum_{\vec{q}'} e^{-(\vec{q}-\vec{q}')^2/q_0^2} {\rm Im}\chi(\vec{q}',\omega) / 
\sum_{\vec{q}'} e^{-(\vec{q}-\vec{q}')^2/q_0^2}$ for the ease of comparison with the experimental data in Fig. \ref{fig:fete-ins}.
For $T>T_{fluc}$, orbital fluctuations do not affect the magnetic excitations. In this case, $H_{OO}$ is completely turned off. For $T_{fluc}>T>T_s$,
the system does not have long-range orbital order but rather has
fluctuating orbital correlations, which can be described at the Gaussian level (in the plot, $\lambda^2=30$ meV
is used). For $T<T_s$, an orbital order is formed and the magnetic excitation is commensurate. The orbital order parameter in the plot is $\phi=20$ meV.}
\end{figure}

It is remarkable to see that our results offer a natural explanation for several puzzles present in previous inelastic neutron scattering measurements.
As shown in Fig. \ref{fig:fete-ins}, for superconducting Fe$_{1-x}$Ni$_x$Te$_{0.5}$Se$_{0.5}$ with $x=0.04$, the overall shape of the magnetic excitations at low energy udergoes 
a distinct transformation from two incommensurate vertical columns to a broad U-shaped spectrum centered at the commensurate wave vectors. 
As shown in Ref. [\onlinecite{xu2012}], the change occurs at $T_{onset} \sim 3T_c$.
Although we can not obtain the temperature-dependence of $\lambda^2$ from a microscopic calculation within the current model, it is expected that 
$\lambda^2$ should gradually increase if the system gets closer to the orbitally ordered state as the temperature is lowered. 
Consequently, this unusual incommensurate-to-commensurate transformation can be captured by our theory, as shown in Fig. \ref{fig:demo}, if we identify $T_{onset}$ as the       
onset temperature of the fluctuating orbital correlations $T_{fluc}$. Moreover, a weak lattice distortion has been found in x-ray measurements\cite{xu2012,gresty2009}, which gives
further support for our orbital-based explanation.

Another interesting observation pointed out by Ewings {\it et. al.}\cite{ewings2011} in inelastic neutron scattering measurement for SrFe$_2$As$_2$ is 
that after a thorough comparison between experimental data and theories, the multiorbital itinerant model treated with RPA\cite{kaneshita2010} 
captures more details of the experimental data compared to a local moment model, except the theory obtained an incommensurate spectrum in the normal state 
which is not seen in experiments. 
This inconsistency can be easily resolved by the inclusion of the Gaussian fluctuation model presented in this paper.  
As seen in Fig. \ref{fig:demo}, while the incommensurate peaks for $T>T_{fluc}$ similar to Ref. [\onlinecite{kaneshita2010}] are reproduced in our calculations, the 
spectral weights of these incommensurate peaks are gradually shifted to $(\pi,0)$ as the static and fluctuating orbital correlations are taken into account.
Another subtle point worthy of mentioning is that there is no soft mode at $\vec{q}=(\pi,\pi)$ in our results, which is an important advantage of itinerant model over 
local moment model as mentioned in Ref. [\onlinecite{ewings2011}].

\section{Discussion}

First, we would like to discuss the nature of the magnetic excitations in spin-based scenarios
proposed for iron-based superconductors.
For the local spin scenario with frustrating $J_1$-$J_2$ interactions, it is generally hard to obtain peaks at incommensurate wave vector in the spin excitation spectrum 
in the paramagnetic state. 
One possibility is to include more longer-range spin interactions
(e.g., $J_3$)\cite{fang2009}.  Even so, it would be difficult to understand why the system would start from an 
incommensurate spin excitation spectrum at high temperature and then
evolve into a stripe-like AFM with a commensurate wave vector.
For the SDW scenario, there are only incommensurate magnetic excitations in the high temperature normal state, 
as seen in Ref. [\onlinecite{xu2012}], and hence this model is not applicable.
As a result, we conclude that the incommensurate-to-commensurate transformation in the magnetic excitations is a unique signature favoring orbital-based 
over spin-based scenarios.  
However, we also recognize the fact that the size of the instantaneous moments is too large to be described by a purely itinerant model in all current existing neutron 
scattering data.
This suggests that one needs to take into account the orbital correlations emerging from an itinerant model to obtain the spin interactions correctly.
Additional inelastic neutron scattering measurements in the high temperature normal state are certainly necessary to resolve the 
long-standing controversy between the orbital-based  and spin-based scenarios.

Second, we comment briefly on materials. 
FeSe is a known iron-based supercondcutor exhibiting only a structural transition and
no magnetic transition\cite{hsu2008,mcqueen2009} and hence is a promising prototype for the orbital-based scenario as well as the incommensurate-to-commensurate 
transformation in its magnetic excitation spectrum. 
The families of iron-based superconductors whose structural and magnetic transition temperatures are well-separated, 
including LaOFeAs\cite{luetkens2009}, CeOFeAs\cite{zhao2009}, NaFeAs\cite{parker2010,wright2012,yi2011,zhang2012}, etc., should also be good candidates to realize the 
physics outlined in this paper.
There have been neutron scattering measurements for parent compounds
of 122, for example, BaFe$_2$As$_2$\cite{harriger2011}, CaFe$_2$As$_2$\cite{zhao20092,diallo2010}, and 
SrFe$_2$As$_2$\cite{ewings2011}, and interestingly they show very different trends. While the magnetic excitations in CaFe$_2$As$_2$ can be well-described by the 
local spin scenario both in the magnetic and the normal states, itinerant models seems to work better in BaFe$_2$As$_2$ and SrFe$_2$As$_2$.
As suggested in Ref. [\onlinecite{ewings2011}], the reason why the local spin scenario fails in SrFe$_2$As$_2$ is that in the high temperature normal state, 
the magnetic excitations at momentum $(\pi,\pi)$ are pushed to high energy which clearly contradicts the prediction of the local spin scenario unless one accepts 
that the anisotropic spin interactions can still be present even above the magnetic transitions.
The most serious problem with the itinerant model for SrFe$_2$As$_2$ is the
prediction of an incommensurate excitation spectrum in the high
temperature normal state.  This can be resolved with the Gaussian
fluctuating orbital correlations proposed here. 
In other words, in order to explain the neutron scattering data of SrFe$_2$As$_2$, the orbital degrees of freedom are necessary even in a spin-based scenario.
Therefore, since the previous work on BaFe$_2$As$_2$ only measured up to 150K, slightly above the magnetic transition temperature ($\sim$ 140K), 
it would be essential to perform additional neutron scattering measurement at much higher temperature to see whether the features observed in 
SrFe$_2$As$_2$ appear in BaFe$_2$As$_2$.  Should they be observed,
this would further substantiate the orbital-based physics proposed here.

Third, we discuss the implication of our theory for superconductivity. 
Since most iron-based superconductors require doping away from the parent compounds to induce superconductivity, 
this strongly suggests that the Fermi surfaces of superconducting samples generally do not have perfect nesting at $(\pi,0)$ and $(0,\pi)$.
Consequently, if the pairing mechanism of iron-based superconductors is really through spin fluctuations\cite{kuroki2008,graser2009}, the necessary 
commensurate spin excitation in this mechanism can not be obtained
without the fluctuating orbital correlations discussed in this paper.
This implies that the role of the orbital degrees of freedom in the superconductivity in iron-based superconductors might be very profound.
Recently, orbital-dependent superconducting gaps have been observed in the superconducting Ba$_{1-x}$K$_x$Fe$_2$As$_2$ via Laser ARPES measurement\cite{malaeb2012}. 
The electronic structure of the vortex core in the FeSe superconductors shows a strong anisotropy\cite{song2011}, which has been interpreted as a 
consequence of competition between nematicity (orbital order) and superconductivity\cite{Chowdhury2011,hung2012}.
A very recent ARPES measurement\cite{sudayama2012} revealed that the doping dependence of the band renormalizations of d$_{xz}$ and d$_{yz}$ orbitals 
is strongly correlated with the enhancement of the superconductivity.
These findings strongly suggest that only an inceptive
understanding of superconductivity would be obtained without including the
orbital degree of freedom.
We propose that the incommensurate-to-commensurate transformation in the magnetic excitations should be a general feature in most superconducting iron pnictides.

Finally, we would like to discuss the magnetic excitation along the direction longitudinal to $Q_{AFM}$. 
Our model showed that the magnetic excitation still disperses along this direction, despite the fact that the spectral weight could be different 
from that along the transverse direction. 
Experimentally, it has been found that the magnetic dispersion is absent along the longitudinal direction for FeTe/Se systems\cite{leesh2010,xu2012}, 
which remains a puzzle and is beyond our current model. 
One possibility is to exploit an effective degenerate double exchange model\cite{lv2010} whose spin interactions are orbital dependent so that the effects of the 
orbital correlations discussed in this paper could be correctly incorporated.
This work is currently under development.
\section{Conclusion}
In this paper, we have studied  magnetic excitations in the normal state of iron based superconductors. 
Because the orbitals of $d_{xz}$ and $d_{yz}$ are closely related to
the shape of the Fermi surface, the orbital order, which distorts the
Fermi surfaces, can have a profound effect on the magnetic excitations. 
We do not specify the origin of the orbital ordering and fluctuations since the results are generally valid regardless of the origin.
Furthermore, we assume that the ferro-orbital fluctuations are the dominating ones as the structure phase transition is approached from higher temperature.
We have shown that orbital order always enhances one of the commensurate spin excitation wave vectors even when the original Fermi surfaces do not have a nesting 
at these commensurate wave vectors.
Furthermore, even when long-range orbital order is absent, Gaussian
fluctuations of the orbital order can still couple to the magnetic
excitations.   We found that these Gaussian fluctuations also tend to diffuse the spectral weights toward the commensurate wave vector, resulting in a 
broad spectrum near the commensurate wave vectors as seen in neutron
scattering measurements\cite{xu2012}.
We have shown that this orbital-based scenario offers a natural explanation for the incommensurate-to-commensurate transformation observed in 
Fe$_{1-x}$Ni$_x$Te$_{0.5}$Se$_{0.5}$, which can hardly be explained by spin-based scenarios. 
We propose that this unusual transformation is an important signature
that can distinguish orbital from spin-based physics in pnictides.
Our theory can be tested by additional neutron scattering experiments at
higher temperatures.

\section{Acknowledgments} 
We would like to thank Andriy Nevidomskyy, Allan H. MacDonald, and Qimiao Si for helpful discussions. 
This work is supported by the Center for Emergent Superconductivity, a DOE Energy Frontier Research Center, Grant No.~DE-AC0298CH1088.
In addition, Weicheng Lv and P. Phillips received research support from the NSF-DMR-1104909.

\end{document}